\documentclass[prb,twocolumn,amsmath,amssymb,floatfix,footinbib,bibnotes]{revtex4-1}

\pdfoutput=1

\usepackage[colorlinks=true,citecolor=blue,linkcolor=magenta]{hyperref}
\usepackage{amsmath}
\usepackage{graphicx}
\usepackage{dsfont}
\usepackage{changepage}
\usepackage{fancyhdr}
\usepackage{amsthm, amssymb}
\usepackage{array}
\usepackage[usenames,dvipsnames]{color}

\newcommand{\be}{\begin{align}}
\newcommand{\ee}{\end{align}}

\def \be{\begin{equation}}
\def \ee{\end{equation}}
\def \ba{\begin{array}}
\def \ea{\end{array}}
\def \bea{\begin{eqnarray}}
\def \eea{\end{eqnarray}}
\def \nn{\nonumber}
\def \e{{\epsilon}}

\def \t{{\theta}}

\def \g{{\gamma}}

\def \e{{\epsilon}}

\def \av#1{{\langle#1\rangle}}

\def \ba{\begin{align*}}
\def \ea{\end{align*}}

\newcounter{indice}

\def \mrm{\mathrm}

\def \mc{\mathcal}

\begin{document}

\title{Topological degeneracy and pairing in a one-dimensional gas of spinless Fermions   }

\author{Jonathan Ruhman$^1$ and Ehud Altman$^2$ \\
{\small \em $^1$Department of Physics, Massachusetts Institute of Technology, Cambridge, MA 02139 USA \\
$^2$Department of Physics, University of California, Berkeley, CA 94720 USA }}
\begin{abstract}
We revisit the low energy physics of one dimensional spinless fermion liquids, showing that with sufficiently strong interactions the conventional Luttinger liquid can give way to a strong pairing phase. While the density fluctuations in both phases are described by a gapless Luttinger liquid, single fermion excitations are gapped only in the strong pairing phase.
Smooth spatial Interfaces between the two phases lead to topological degeneracies in the ground state and low energy phonon spectrum.
 Using a concrete microscopic model, with both single particle and pair hopping, we show that the strong pairing state is established through emergence of a new low energy fermionic mode. We characterize the two phases with numerical  calculations using the density matrix renormalization group. In particular we find enhancement of the central charge from $c=1$ in the two Luttinger liquid phases to $c=3/2$ at the critical point, which gives direct evidence for an emergent critical Majorana mode. Finally, we confirm the existence of topological degeneracies in the low energy phonon spectrum, associated with spatial interfaces between the two phases.
\end{abstract}
\maketitle

\section{Introduction}
A one dimensional superconductor of spinless fermions features a topological phase and phase transition, as first noted  by Kitaev \cite{Kitaev2001}.
In such a system, changing the ratio between a mean field pairing potential and the chemical potential tunes a transition from a topological "weak pairing" phase to a trivial "strong pairing" phase. Majorana zero modes occur at spatial interfaces between these two states or at the boundary of the topological phase with vacuum. Such a mean field picture is, however, valid in a one dimensional system only if the pairing field is imposed by proximity coupling to an external higher dimensional superconductor, which explicitly breaks the $U(1)$ charge symmetry. It is natural to ask if analogous phases and phase transitions can occur in a charge conserving strictly one dimensional system of spinless fermions, which is necessarily gapless.

Here we address this question using an effective low energy theory as well as DMRG calculations of a microscopic lattice model. Intuitively, one might think that a transition from weak to strong pairing can be driven by increasing the attractive interactions between fermions. For weak interactions the system should form a Luttinger liquid with power-law decay of the pairing correlations of both the single particle and pairing correlations. Single fermion excitations are gapless. For strong attractive interactions, on the other hand, one can imagine formation of a liquid of molecules made of strongly bound fermion pairs with a gap to single fermion excitations. A technical problem, that simple nearest neighbor interactions lead to clustering of particles and consequent phase separation, may be resolved by modifying the interaction potential. There is however a more subtle difficulty concerning such a transition from strong to weak pairing, which arises when considering the  low energy theory of spinless fermions.

The standard long wavelength description of spinless fermions, obtained by bosonization, leads to a Luttinger liquid with gapless single particle excitations \cite{giamarchi2004quantum}. On the other hand, the strong pairing phase must feature two distinct modes, a gapless phonon (charge) mode and another mode reflecting the gapped single particle excitations. If there is a continuous transition between these two states, then there should be a way to include the gapped single particle mode
within a low energy theory. It is not clear a priori  how this can be done when microscopically we have a single mode of spinless fermions.

A single particle gap is easier to establish in systems with spin. Indeed, previous work on such systems with charge conservation focused on models of interacting spin-1/2 electrons with spin orbit coupling and a Zeeman field \cite{Fidkowski2011,Sau2011,Ruhman2015}. The electron spin contributes a degree of freedom with a gap that can be tuned across a quantum phase transition at which the gap vanishes and changes its character. The phase on one side of the transition is adiabatically connected to the limit of a large Zeeman field where the electrons are almost polarized. Hence this phase is identical to the spinless Luttinger liquid, the weak pairing phase discussed above. The other phase is adiabatically connected to the limit of vanishing Zeeman field, where electrons pair up to form a spin-gapped Luttinger liquid. This state, with a gap to single electron excitations, is analogous to the strong pairing phase. The two phases are separated by a quantum critical point with central charge $3/2$ and Majorana-like zero modes occur at spatial boundaries between the two phases.

In this paper we demonstrate that the same phases and phase transition occur also in the case of spinless fermions. We show that the additional degree of freedom required to generate the gap to single fermion excitations arises from an emergent mode. While our paper was in writing we became aware of a related work \cite{kane2017pairing}, in which  the emergence of a fermion mode in the transition between strong and weak pairing of spinless fermions was postulated. Here we  show how this mode arises in a microscopic description, from which we can explicitly derive the effective theory with the emergent low energy mode. We then demonstrate the existence of the two phases, as well as the quantum critical point with central charge $3/2$ that separates them, using DMRG simulations~\cite{White1992,ITensor}. We show that in another part of the parameter space the two phases are separated by a 1st order transition. Finally, we present numerical evidence for the Majorana like zero modes bound to interfaces between the two phases.

\begin{figure}%
   \includegraphics[width=0.9\linewidth]{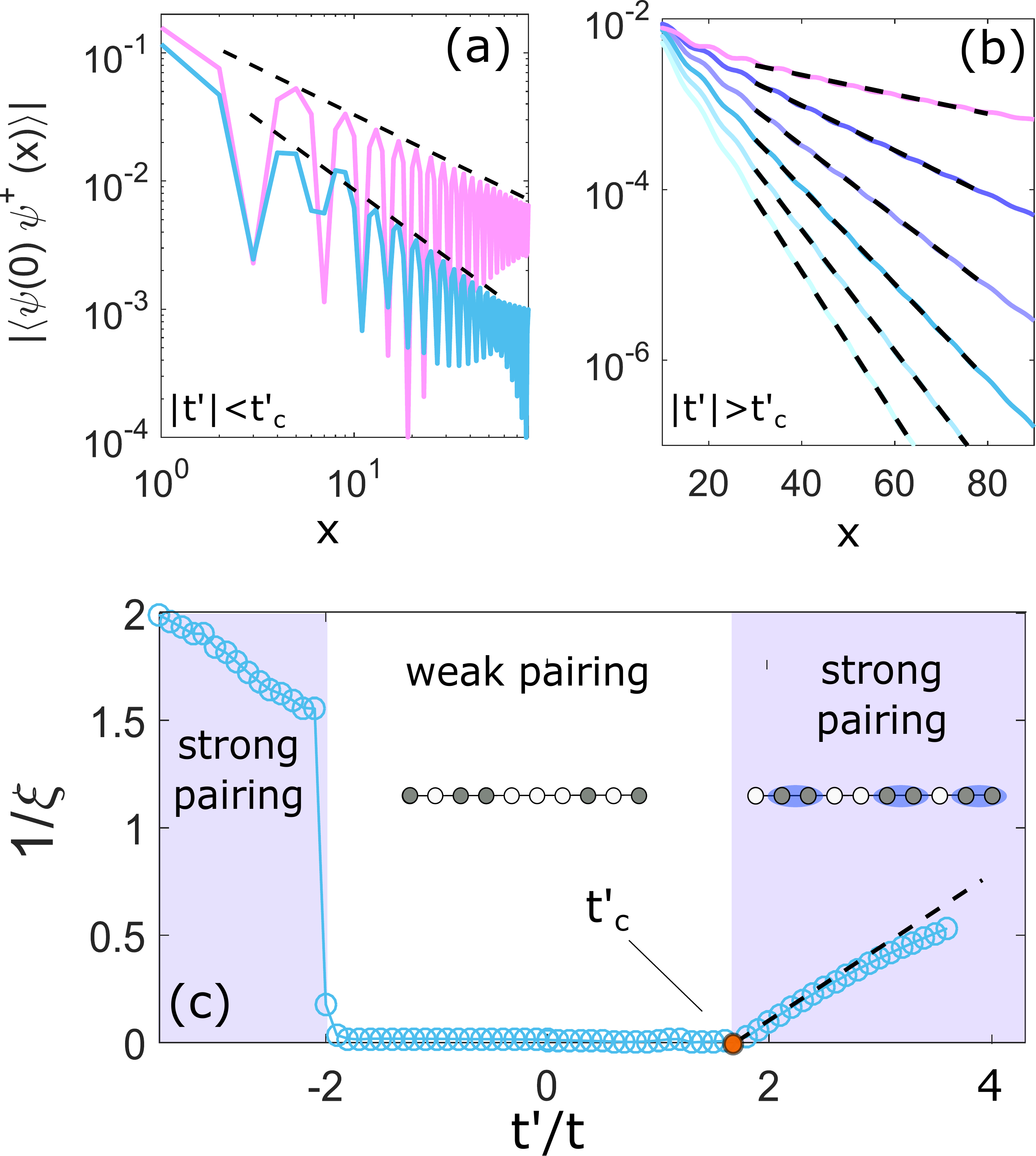}
   \caption{(a) The single particle correlation function in the weak-pairing phase for $t' = 1.3$ (blue) and $t' = 0.7$ (pink) obtained from the DMRG simulation with $V = 0.3$, $t = 1$, $n = 1/4$ and $N = 184$.
   We find that in this phase the single particle correlations exhibit power law decay (Dashed lines indicate a power law). (b) The same correlation function in the strong pairing phase ($t'  = 1.8,1.9,\ldots,2.3$), where we observe exponential decay. Dashed lines indicate the exponential fits. (c) The inverse decay length, $\xi^{-1}$, of the single particle correlation function $|\langle \psi_0  \psi^\dag _x \rangle|$ vs. $t'$ inferred from panel (b). We find two transitions into a paired state. For positive  $t'$ the transition is continuous with $1/\xi \propto |t'-t'_c|$ (dashed line) consistent with an Ising critical point. For negative $t'$ the transition is strongly first order.
%    The inset  depicts the two phases : "weak pairing" - a single mode Luttinger liquid and "strong pairing" a Luttinger liquid of bosonic pairs.
    }\label{fig:xi}
\end{figure}

\begin{figure}%
\centering
  \includegraphics[width=1\linewidth]{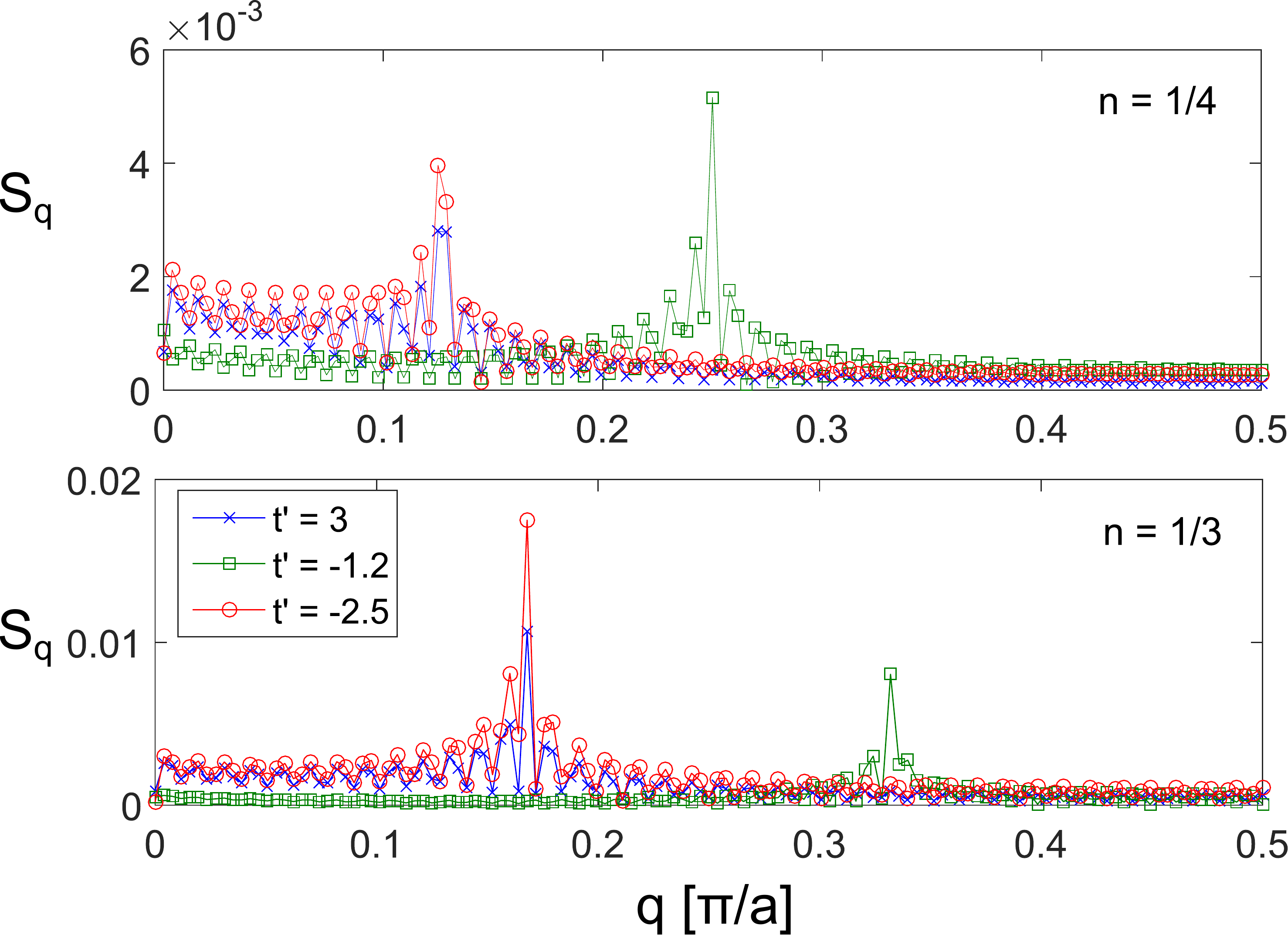} %
    \caption{The static structure factor (Fourier transform of the density-density correlation function) for three values of the pair hopping $t'= -2.5,-1.2,\mrm{and}\,3$ (red circles, green squares and blue crosses, respectively) and for two different filling factors $n = 1/4$ (top) and $n = 1/3$ (bottom). For $t' = -1.2$ the system is in the weak-pairing phase and the usual $2k_F = \pi n $ peak is observed. On the other hand for $t' = 3$ and $-2.5$ the system is in the strong pairing phase and the peak appears at $\pi n_b = \pi n/2$.  In both panels we used $V = 0.3$ and $N  = 184$. }%
    \label{fig:nn} %
\end{figure}

\section{Model}
Our starting point for theoretical and numerical analysis is the following interacting model of spinless fermions:
\begin{align}
H = &-\sum_{x = 1}^N \bigg[ t\,\psi^\dag _{x+1} \psi_x +t'\,\psi^\dag _{x+1} \psi^\dag_x \psi_x \psi_{x-1}+\mrm{H.c.}\bigg] \nn\\&
+\sum_{x = 1}^N  V\,n_x n_{x+1}\,, \label{H}
\end{align}
where $\psi_x$ and $\psi^\dag_x$ are the creation and annihilation operators of a Fermion at site $x$ and the second term (proportional to $t'$) can be viewed as hopping of a pair from the bond at $x-1/2$ to $x+1/2$. The main reason for introducing this model rather than considering the simplest interacting model is to avoid phase separation. A pair hopping term favors pairing as it gains from kinetic energy of pairs. A simple nearest neighbor interaction on the other hand would benefit from clustering of many particles more than from pairing, leading to phase separation. Another advantage of this model, as we show below, is that it allows for a simple derivation of a low energy effective theory that includes the additional mode required to generate the single particle gap.

\section{Mean-field approximation}
As mentioned, in the limit of weak coupling ($t',\,V\ll t$) the low-energy behavior of the system is captured by a single mode Luttinger liquid obtained from straightforwardly bosonizing the Fermionic operators~\cite{giamarchi2004quantum} in Eq.~(\ref{H}). However, in the limit of strong coupling, where the pair hopping term $t'$ is large, the bosonization approach breaks down. Nonetheless, as we show now, it is very easy to obtain the effective low-energy theory describing the strong coupling regime starting from Eq.~(\ref{H}).

To show this we use a simple mean field  approximation to decouple the pair hopping term and the interaction term $V$. First write the density as $\psi^\dag_x\psi_x = n+\delta n_x$ and the operator $\psi^\dag_{x+1}\psi_{x-1}+\mrm{H.c.}=\chi+\delta\chi_x$. Here $n$ is the average density and $\chi=\av{\psi^\dag_{x+1}\psi_{x-1}+\mrm{H.c.}}$.
Plugging this in Eq.~(\ref{H}) and neglecting second order in the fluctuations we obtain
\begin{align}
H =-\sum_{x=1}^N \bigg[t\, \psi^\dag _{x+1}\psi_x +n\,t'\,\, \psi^\dag _{x+1} \psi_{x-1} +\mrm{H.c.}\bigg],\label{H2}
\nn
\end{align}
where we have absorbed terms linear in the density into a constant chemical potential.  Thus, within this mean field theory the pair hopping term effectively generates next-nearest neighbor hopping. The resulting single-particle dispersion is given by $\e_k = -2 t \cos k -2n\,t'\, \cos 2k$.
The fermion modes that emerge from the  from the low energy spectrum of the mean field Hamiltonian can serve as the basis for a low energy effective theory. The most relevant low energy components of the neglected fluctuations terms can then be reintroduced to this theory.

\begin{figure}%
\centering
  \includegraphics[width=1\linewidth]{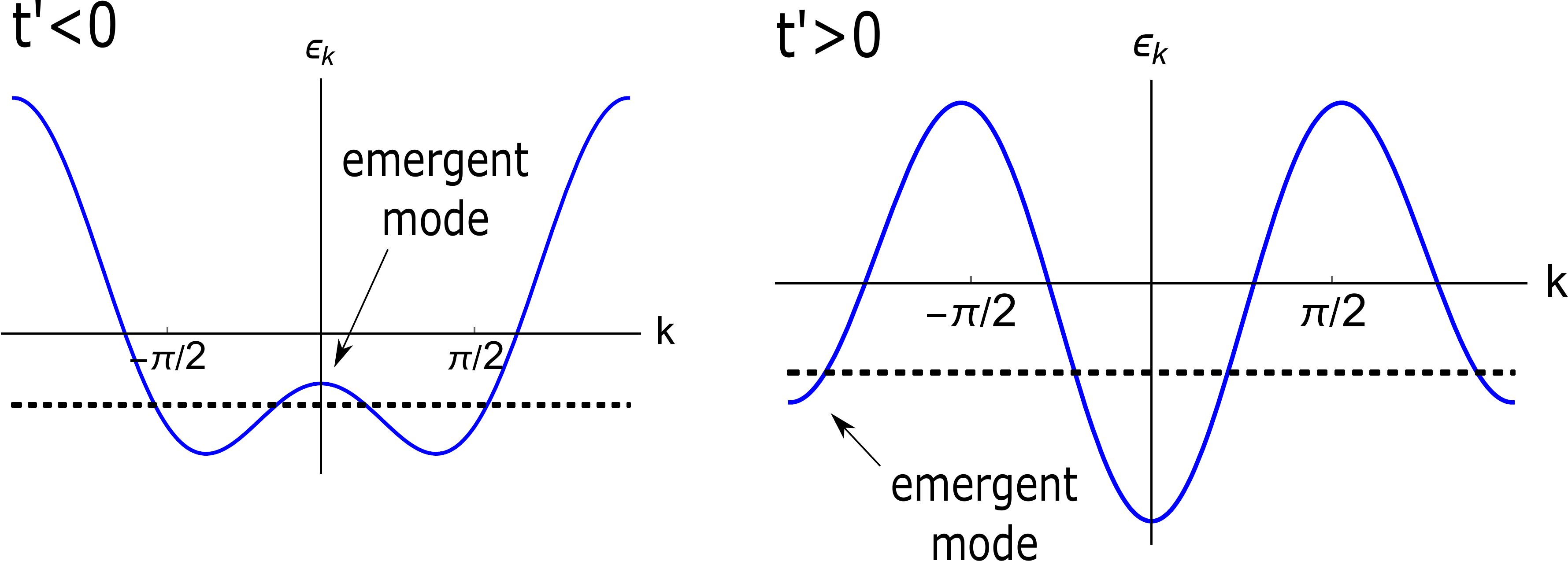} %
    \caption{ Schematic plot of the effective dispersion $\e_k$ for large $|t'|$ (beyond the critical value). The dashed line is the Fermi energy. The left panel corresponds to $t'<0$. There the emergent mode crosses the Fermi energy near $k = 0$. The right panel corresponds to $t'>0$, where the new mode appears at $k = \pi$.  }%
    \label{fig:dispersion} %
\end{figure}

Fig.~\ref{fig:xi}.(a) shows the mean field dispersion obtained for positive and negative values of $t'$, where the dashed black line is at the Fermi energy. In both cases a  non chiral fermion mode  approaches the Fermi energy and crosses it when $|t'|$ exceeds a certain threshold. Note the perfect analogy with the low energy mode structure in the spinfull Fermi system with spin-orbit coupling and a Zeeman field \cite{Ruhman2015}. In that system the second mode approaches the Fermi energy near $k=0$ as the Zeeman field is reduced below threshold, leading to a topological transition into a state with a gap to single fermion excitations. In the spinless system we consider here the new modes  appear near $k = 0$ for $t'<0$ and near $k = \pi$ for $t'>0$, as shown in Fig. \ref{fig:dispersion}.

\section{Effective field theory} We now review the low energy theory used to describe a new fermion mode approaching the pre-existing Fermi surface \cite{Meyer2007,Sitte2009,Ruhman2012,Ruhman2015,alberton2016fate}. Assuming for now that the new mode has crossed the Fermi surface, we can express the fermion creation operator in terms of four slow modes:
\bea
\psi(x) &\approx& e^{ik_F x} R_0(x) +e^{-ik_F x} L_0(x) \nn\\
&&+e^{ik_1 x} R_1(x)+e^{-ik_1 x} L_1(x) \,,\label{psi} \eea
Where $L_n,R_n$ are the left and right moving modes and $n=0,1$ denote respectively the pre-existing modes at $\pm k_F$ and the new low energy modes at $k_1$ ($k_1 = \pi,0$ for $t'>0$ and $t'<0$ respectively).
The situation in which the new mode is still gapped is incorporated in the low energy theory through the coupling $- V_1=\int dx  L_1^\dagger R_1 +\text{H.c.}$, which gives a mass to this mode.

The long wavelength fermion modes are bosonized using the standard identities $R_n\sim {1 \over  \sqrt{2\pi  a}} \exp\left[ i\t_n -i\phi_n \right]$ and $L_n \sim {1 \over  \sqrt{2\pi  a}} \exp\left[ i\t_n +i\phi_n  \right]$. Here $a$ is the lattice constant, the partial charge density carried by mode $n$ is $\partial_x \phi_n$ and the current $\partial_x \t_n$, so that $[ \partial_x,\phi(x),\t(x')] = -i \pi \delta (x-x')$. In this bosonized theory  the linearly dispersing fermion modes and the forward scattering components of the quartic interactions lead, as usual, to two independent harmonic theories (Luttinger liquids). The mass gap of the new fermion mode on the other hand translates to the cosine term $V_1= -\tilde{g}_1\int dx \cos(2\phi_1)$. Finally, there is a back-scattering contribution $-g_{int} \int dx L_1^\dagger R_1^\dagger R_2 L_2$ coming from the quartic couplings (pair hopping) in the microscopic hamiltonian, which translates into another cosine term  $V_{int}=-g_{int}\int dx \cos[2(\theta_1-\theta_0)]$.

The last cosine term couples the two Luttinger liquid modes. This situation can be simplified by the following canonical transformation
\[\begin{matrix} \;\;\;\;\;\;\;\phi_+ = \phi_0 + \phi_1 \\ \t_+ = \t_0\;\;\; \end{matrix}\;\;\;\begin{matrix} \phi_- = \phi_1\\ \;\;\;\;\;\;\;\t_- = \t_1 - \t_0 \end{matrix}\,,\]
In this representation the coupling through the cosine terms are replaced by linear coupling between the two modes $\partial_x \t_- \partial_x \t_+$ and $\partial_x \phi_- \partial_x \phi_+$. However, as we'll see shortly these couplings are irrelevant in both phases.
Ignoring the linear coupling terms for now, the Hamiltonian takes the form $\mc H = \mc H_+  + \mc H_- $, with
\begin{align}
\mc H_+={u_+\over 2\pi}\int dx &\left[ {{K_{+}}(\partial_x \t_+)^2+{1\over K_+}(\partial_x \phi_+)^2 }\right]\label{H+} \\
\mc H_-={u_-\over 2\pi}\int dx &\left[ {{K_{-}}(\partial_x \t_-)^2+{1\over K_-}(\partial_x \phi_-)^2 }\right] \label{H-}\\
-&\int dx \left[ g_1\, \cos 2\phi_- + g_i \, \cos 2\t_- \right],\nn
\end{align}

Both cosine terms shown above are relevant perturbations in the hamiltonian $H_-$ leading to two distinct phases with a transition tuned by the ratio $g_1/g_i$. If the coupling $g_1$ is dominant it pins  $\phi_-$, such that  $\av{\phi_-}=\av{\phi_0}=0$, while the phase $\t_-$ is strongly fluctuating. This is the phase established when the new fermion mode approaching the Fermi surface is still gapped by the quadratic back scattering. Hence this corresponds to the weak pairing phase, adiabatically connected to the usual  Luttinger liquid of spinless Fermions. On the other hand, when the interaction $g_i$ dominates it pins the field $\t_-$ while $\phi_-$ is fluctuating. This is the strong pairing phase with a gap to single fermion excitations. We term the Hamiltonian $H_-$ the parity sector of the theory, because the phase realized in this Hamiltonian determines whether the system has a gap to excitations that change fermion parity. In either phase the linear coupling terms discussed above are irrelevant.

The fundamental distinction between the two phases manifests in the decay of the single fermion correlation function $\av{\psi^\dagger(x)\psi(0)}$. The long distance behavior of this correlation can be computed by expressing $\psi(x)$ in Eq. (\ref{psi}) using the Bose fields  $\t_\pm(x)$ and $\phi_\pm(x)$. In the weak pairing phase where $\phi_-$ is pinned, contributions involving $e^{i\phi_+}$ can be set to a constant, while contributions from the strongly fluctuating operator $e^{i\t_-}$ can be neglected in the long wavelength limit. Thus  the leading long-wavelength contribution to the fermion operator in this phase is
\be
\psi(x)= {1\over \sqrt{2\pi a}} e^{ik_F x}e^{i\t_+ - i\phi_+}+{1\over \sqrt{2\pi a}} e^{-i k_F x}e^{i\t_+ + i\phi_+},
\ee
exactly as in a conventional Luttinger liquid. We get the single particle correlation function
%where $r/\pi = g_\phi - g_\t$, $\partial_x \phi_I = -\pi \tilde \psi^\dag_e \tilde \psi $ and using Eq.~\ref{transformation} we get that $\t_I = \t - \t_\psi$ where $\t_\psi$ is the phase of the emergent mode $\psi_e$. In terms of these fields the original modes are given by\cite{Ruhman2012}
%\begin{align}
%& R(x),L(x) \sim {e^{\pm i k_F x}\over \sqrt{2\pi a} } e^{\mp i \tilde \phi + i\t} e^{\pm i\phi_I} \label{fields}\\
%&\psi_e(x) \sim  {e^{ i k_F x+i{3\pi\over 4}}\over \sqrt{2\pi a} } \left[ \cos \left(\t_I - \phi_I \right)+i \sin\left(\t_I + \phi_I \right) \right] e^{-i \t}\nn
%\end{align}
\[ \langle \psi(0) \psi(x) \rangle \propto  \cos(2 k_Fx) \left({a\over |x| } \right)^{{1\over K_+} + K_+}\,.\]

In the strong pairing phase $\t_-$ is pinned while  $e^{i\phi_-}$ has exponentially decaying correlations. Since all the modes $R_n$ and $L_n$ have a contribution from $e^{i\phi_+}$, the single particle correlation function also decays exponentially:
\[ \langle \psi(0) \psi(x) \rangle \propto \exp\left({- |x| / \xi }\right)\,.\]

The sharp change in the behavior of the single particle correlation function in the two phases is a direct consequence of having a gap to fermion excitations (i.e. a parity gap) in the strong pairing phase and no such gap in the weak pairing phase.
Below we look for this signature of the two phases in the DMRG calculations of the microscopic model (\ref{H}).

The  transition between the weak and strong pairing phases, implied by the effective field theory (\ref{H-}) of the parity sector, is of the Ising universality class \cite{Lecheminant2002}. To a first approximation the parity sector is decoupled from the charge sector, which always forms a gapless Luttinger liquid. Couplings of the two sectors is irrelevant in certain regimes, leading only to logarithmic corrections to the Ising criticality, while in a different regime they lead to a weak first order transition \cite{Sitte2009,alberton2016fate}.

The emergence of a critical Ising mode implies an enhanced central charge at the critical point.
On either side  of the transition, which implies a central charge $c = 1$ due to the gapless charge sector (\ref{H+}). At the transition point, however, the parity sector becomes gapless and contributes an additional $1/2 $, leading to an overall central charge of $c=3/2$. The appearance of central charge greater than unity is direct evidence of an emergent mode, since our starting point was an interacting model of spinless Fermions, which is naively expected to have the low energy behavior of a single mode Luttinger liquid with $c = 1$.

\section{Numerics}
We now turn to the numerical simulations. We obtain the ground state of the microscopic Hamiltonian (\ref{H}) on open boundary conditions using the single-site DMRG algorithm~\cite{White1992} provided by ITensor~\cite{ITensor}. In these simulations we take $n = 1/4$, $V = 0.3t$ and $N = 186$, unless indicated otherwise. The pair hopping, $t'$, is used as the tuning parameter of the weak to strong pairing transition.

We start by analyzing the single particle correlation function $\av{\psi^\dagger(x)\psi(0)}$, which is expected to decay as a power law at long distances in the weak pairing phase and exponentially in the strong pairing state, due to the gap in the single fermion spectrum. This correlation function is plotted in Fig. ~[\ref{fig:xi}.a,b] for different values of the pair hopping term. For $t'< 1.8$ (Fig. ~[\ref{fig:xi}.a]) the correlation function exhibits $2k_F$ oscillations with a power law decaying envelope. On the other hand, in Fig. ~[\ref{fig:xi}.b], we show that for the larger values of the pair hopping, $t'>1.8$, the correlations decay exponentially indicating strong pairing.
The range of $t'$, where strong pairing is observed is indicated in Fig. ~[\ref{fig:xi}.c] by the purple shading (the white region marks the weak pairing phase).

The density-density correlations give further insight into the nature of the two phases. In both cases the oscillating part of the correlation decays as a non universal power law (related to the Luttinger parameter), however the period of the oscillation is doubled in the strong pairing state.
This is clearly seen in Fig.~\ref{fig:nn} showing the static structure factors (Fourier transforms of the density correlations) for two different filling factors $n = 1/4$ and $n = 1/3$ and different values of $t'$.
For values of $t'$ corresponding to the weak pairing phase in Fig.~[\ref{fig:xi}.c] there is a clear peak at $\pi \, n$ reflecting the fermion density. On the other hand, for values of $t'$ corresponding to the strong pairing phase the peak frequency is at $\pi\,n/2$. This period reflects the structure factor associated with a liquid of fermion pairs with boson density $n/2$.
We also note that the pair-pair correlation function exhibits power law decay in both phases (see Appendix. A).

Having characterized the two phases we turn our attention to the phase transition separating them. In particular Fig.~[\ref{fig:xi}.c] shows the behavior of the inverse correlation length   $1/\xi$ associated with the decay of the single particle correlation function. In the weak pairing phase where these correlations exhibit a power-law decay $1/\xi=0$, while $1/\xi>0$ in the strong pairing phase. The data suggests a continuous phase transition at a critical value $t_c'>0$ at which $1/\xi$ vanishes.
%Moreover, the singularity of the correlation length is consistent with $1/\xi\sim (t-t_c)^{\nu}$ with $\nu=1$ as expected in an Ising critical point.
On the other hand, the transition at the negative value of pair hopping $t_c'<0$ shows a large jump in $1/\xi$ indicating a first order phase transition at this point.

The mean field theory discussed above suggests that the difference between the two transitions may stem from the very different dispersions associated with the emergent mode in the two cases. As seen in Fig. \ref{fig:dispersion},  the Fermi velocity associated with this mode near $t_c'<0$ is much smaller than the bare Fermi velocity. The large effective mass therefore makes this state highly susceptible to phase separation. On the other hand for $t'>0$ the Fermi velocity is higher than its bare value reducing the susceptibility to phase separation.

Let us discuss the critical properties of the  continuous transition at  positive pair hopping $t'_c \approx 1.8$. The vanishing of $1/\xi$ is consistent with being linear in $t'-t'_c$ as expected in an Ising critical point. However a much better indication of the critical behavior is obtained by analyzing the entanglement entropy of a subsystem.  Since the ground state is critical, both at the transition and on either side of it, we expect the entanglement entropy to scale logarithmically with the system size. More precisely, the entanglement entropy of a sub-system of size $x$ in a system of size $N$ with periodic boundary conditions is given by\cite{calabrese2009entanglement}
\be
 S_{\mrm{vN}} = {c \over 3} \log\left[ { {N \over \pi} \sin {\pi x \over N} }\right]+\g\,,  \label{c}
\ee
%{\cred{[This  formula is specific to periodic boundary conditions, right?]}}
where $c$ is the central charge and $\g$ is a constant associated with short range entanglement. Both the strong and weak pairing phases are described by Luttinger liquids with $c=1$ in the long wave-length limit. However at the critical point we expect to observe an enhancement of the central charge to $c = 3/2$ suggesting emergence of a critical Ising degree of freedom in addition to the pre-existing Luttinger liquid.

\begin{figure}%
\centering
  \includegraphics[width=0.9\linewidth]{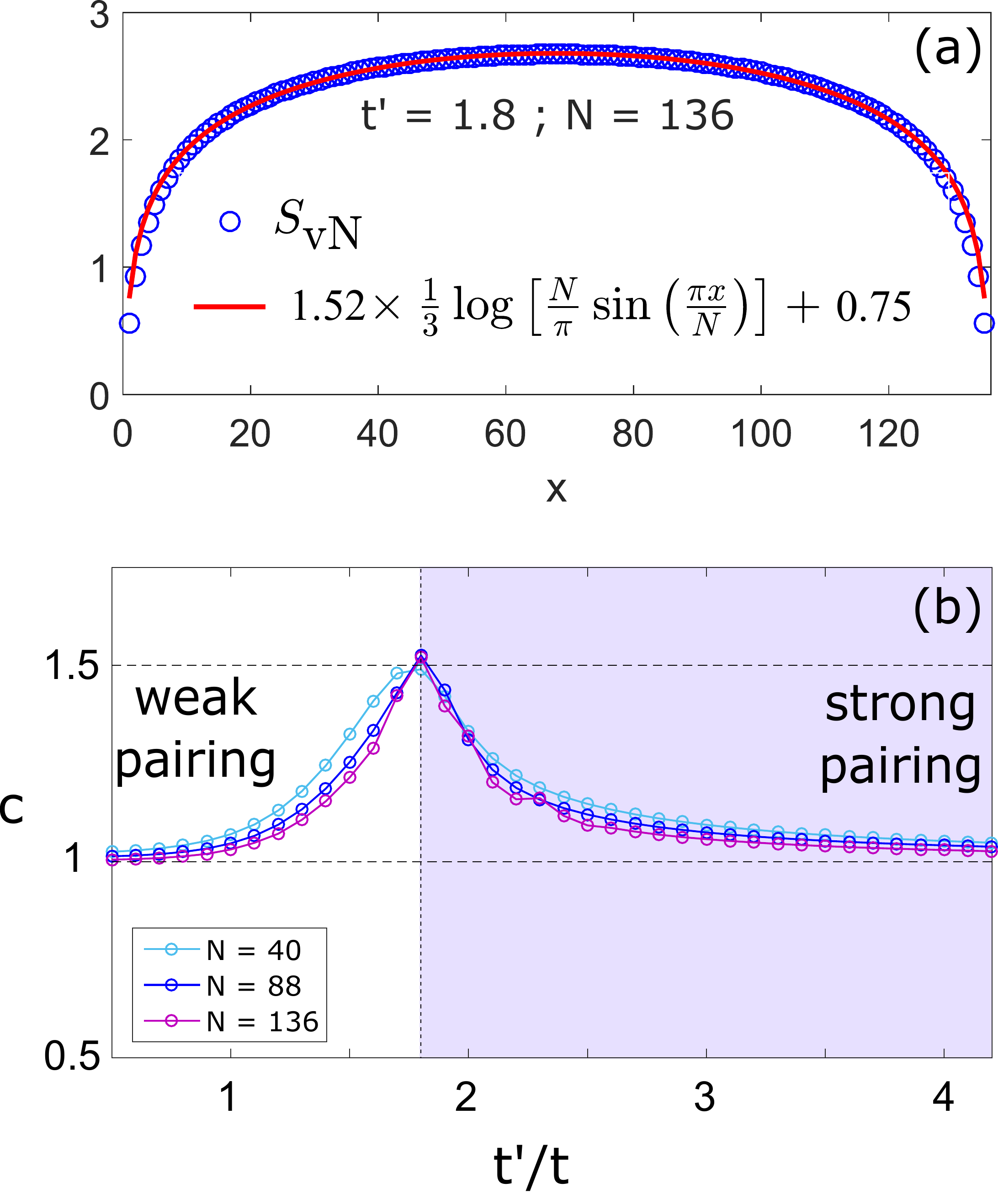} %
    \caption{ (a) The von Neumann entropy near the critical point $t' = 1.8$ (blue circles) and a fit to Eq.~(\ref{c}) (red line) using $c = 1.53$, $\g = 0.755$ and $N = 136$. The rest of the parameters in the simulation are $n = 1/4$ and $V = 0.3t$. (b) The central charge obtained from fitting the entanglement entropy to Eq.~(\ref{c}) as a function of $t'$ for $N = 40,88,136$. Away from the critical point the central charge converges to $c = 1$ and approaches $c = 3/2$ at the critical point $t' = 1.8$.  }%
    \label{fig:central charge} %
\end{figure}

%The periodic boundary conditions are simulated using one long bond connecting the first and last unit cell (where in our case the unit cell size is two due to the pair hopping).
This expectation is  confirmed by our numerical results shown in Fig. \ref{fig:central charge}. We note that these calculations are done with periodic boundary conditions in order to allow a clean fit to the CFT prediction \cite{calabrese2009entanglement} . As an example, in Fig. [\ref{fig:central charge}.a] we plot the fit of the von Neumann entropy (blue circles) to Eq.~(\ref{c}) (red line) for the biggest system size, $N = 136$, at the critical point $t' = 1.8$, using $n = 1/4$, $V = 0.3t$. The best fit is obtained for $c = 1.53$ and $\g = 0.755$. Fig. \ref{fig:central charge}.(b) shows the fitted central charge as a function of $t'$ for three different lengths ($N = 40,\,88$ and $136$). %To obtain this fit we have fixed  $\g$ independent of the length, as required by Eq.~(\ref{c}).
Hence the numerical evidence clearly confirms the emergence of a critical Ising (Majorana) mode at the quantum phase transition between the weak and strong pairing phases of Eq.~(\ref{H}).

\section{Topological ground state degeneracy}
As discussed in the introduction, the phases we identify are the closest analogues, in a charge conserving system, of the topological and trivial p-wave superconducting states. In particular similar degeneracies are expected when the system contains interfaces between the two states. A minimal condition for a degeneracy in the charge conserving system is having at least two regions of the weak pairing phase separated by a region of the strong pairing phase (see refs.~[\onlinecite{Fidkowski2011,Ruhman2015}]).

Let us label states with the quantum numbers $\Pi_R,\Pi_L=\pm 1$ corresponding to the fermion parities in the right and left weak-pairing regions (the parity of the middle region is fixed to $+1$ due to the pairing gap). While the total parity  $\Pi_L+ \Pi_R$ is fixed by the conserved total particle number, the relative parity $\Pi_-=\Pi_L- \Pi_R$ is in-principle undetermined. States with $\Pi_-=\pm 1$ can be connected only by tunneling a single particle through the strong pairing phase in the middle. Hence the off diagonal matrix element is exponentially small in the length of that region. Moreover, for potentials that are smooth on the scale of the system size, the diagonal splitting between the two relative-parity states is also exponentially small \cite{Ruhman2015}. Hence, in this case we expect to observe a near double degeneracy of the ground state, with an energy splitting exponentially small in the separation between the left and right regions, much smaller than the gap to low energy phonon excitations, which scales as $1/N$. Moreover, the argument holds  also for the  low energy phonon excitations (below the parity gap), which are therefore all expected to show a double degeneracy with exponential splitting.

\begin{figure}[h]%
\centering
  \includegraphics[width=0.95\linewidth]{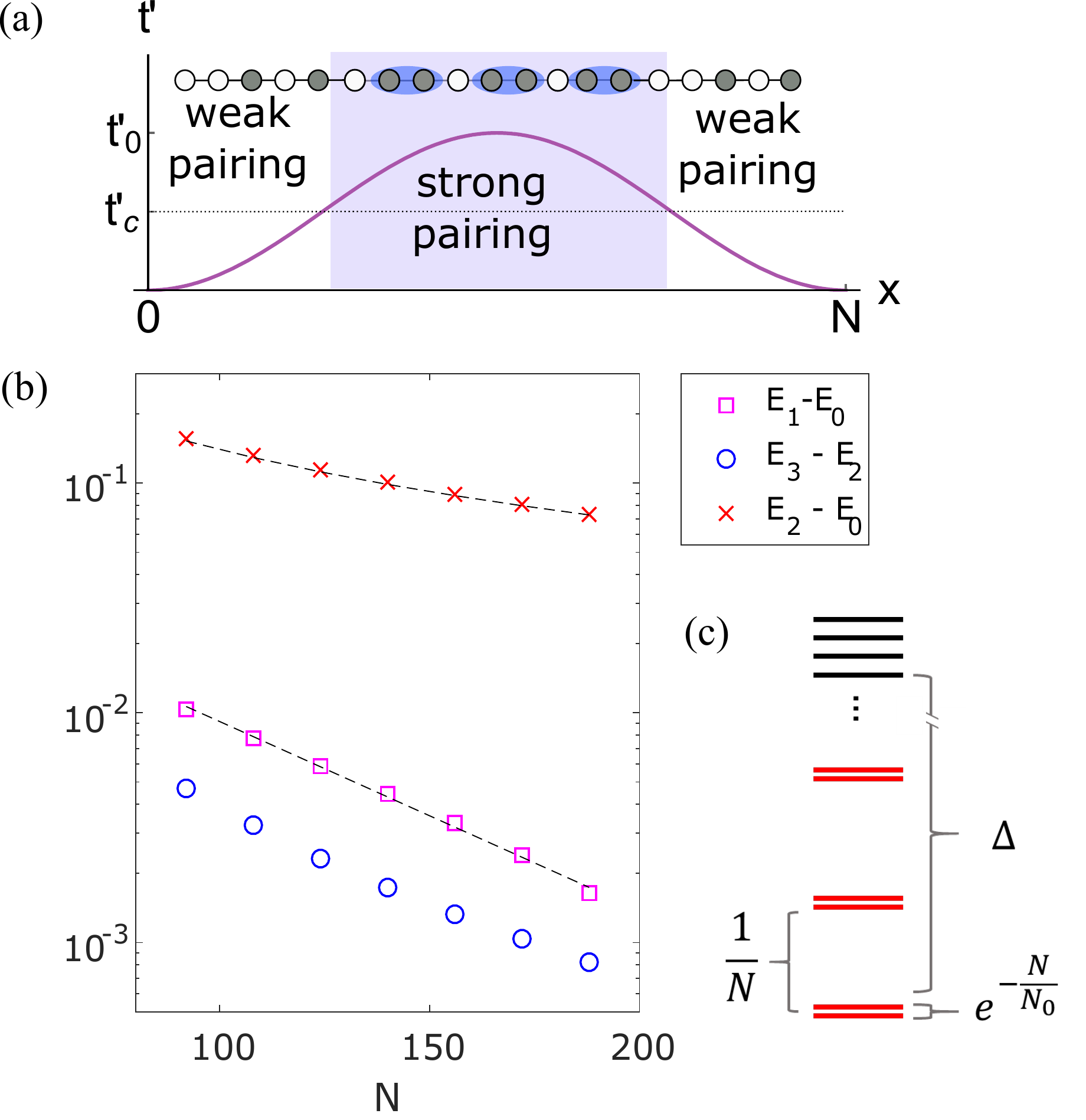} %
    \caption{(a) Inhomogeneous pair hopping $t'(x) = t'_0 \sin^2{\pi x \over N}$ used in the numerical calculations to create a strong pairing region in the center and two weak pairing regions in the wings of the chain. (b) The energy differences between the four lowest states, $E_1-E_0$ (squares), $E_4-E_3$ (circles) and $E_3 - E_0$ (x's) plotted as a function of the system size $N$. The parameters used in the DMRG (ITensor) calculations are $t'_0 = 3.75$, $V = 0.3$ and $n = 1/4$. The dashed lines are fits to $\exp\left( -N/N_0\right)$ for $E_1-E_0$ and $N_0/N$ for $E_2-E_0$.  (c) The expected spectrum exhibits sets of two-fold degenerate states (where the degeneracy scales exponentially with system size $\sim e^{- N/N_0}$) separated by gapes which scale like $\sim1/N$.}%
    \label{fig:gs_degeneracy}%
\end{figure}

To test these predictions we use an inhomogeneous pair hopping $t'(x) = t_0 \sin^2 {\pi x \over N}$ as shown in Fig.~\ref{fig:gs_degeneracy}a.  In this setup $t'(x)>t_c'$ in the middle of the trap, leading to a strong pairing phase in the region $|x|< x_c$, while $t'(x)< t_c'$ in the left and right wings ($|x|>x_c$). We also apply a position dependent potential $\mu(x)-2t'(x)$ in order to set a homogenous fermion density in the trap. Both potentials vary smoothly on the scale of the system size $N$. The calculated energy differences of the four lowest excitations versus system size $N$ are shown in Fig. \ref{fig:gs_degeneracy}b. These differences clearly imply pairing of the energy levels into exponentially split doublets with the general structure illustrated in Fig.  \ref{fig:gs_degeneracy}c.
%
%{\cred{. Finally, .[Yonatan this remark is not clear. Can you elaborate? Do we need to elaborate?]}}

\section{Conclusions}
We have shown that a one dimensional liquid of spinless fermions can undergo a transition between a weak pairing phase and a strong pairing phase. Both phases realize a gapless Luttinger liquid, however single Fermions excitations are gapped only in the strong pairing phase. These two phases are closely related to the topological  and trivial phases of fermions coupled to an external pairing field that explicitly breaks charge conservation.
In order to have a continuous transition between the two phases in a charge conserving system, an additional fermion mode must emerge at low energies. We have demonstrated the mechanism by which a new mode can emerge and derived an effective low energy theory starting from a concrete microscopic model. We have characterized the weak and strong pairing phases as well as the  critical point which separates them using numerical DMRG simulations. Using this approach we investigated an inhomogenous system with two weak-pairing regions separated by a  strong pairing region, which is expected to display topological degeneracies. Specifically, we observed a double degeneracy, with exponentially small splitting in the system size, of the ground state and low energy phonon excitations.
An intriguing question for future study concerns the robustness of quantum memories stored in the relative parity states at temperature well below the parity gap of the middle region but well above the finite size phonon level spacing ($\sim 1/N$).

{\it Acknowledgments -- }
We are grateful to Miles Stoudemire and Anna Keselman for help setting up the DMRG calculation. We especially thank Miles for providing the ITensor\cite{ITensor} package. EA thanks Andrei Bernevig for insightful discussions which helped initiate this study. JR acknowledges the Gordon and Betty Moore Foundation under the EPiQS initiative under grant no. GBMF4303. This work was supported in part by the ERC synergy grant UQUAM.
%
%\emph{\cred{ [Should we say this again here? I have noted it once in the intro.] {Note added ---} During the completion of this paper we have become aware of Ref.~[\onlinecite{kane2017pairing}], which has overlapping results.}}

\appendix

\section{density-density and pair-pair correlations in the strong pairing phase}
\begin{figure}[!htb]%
\centering
  \includegraphics[width=1\linewidth]{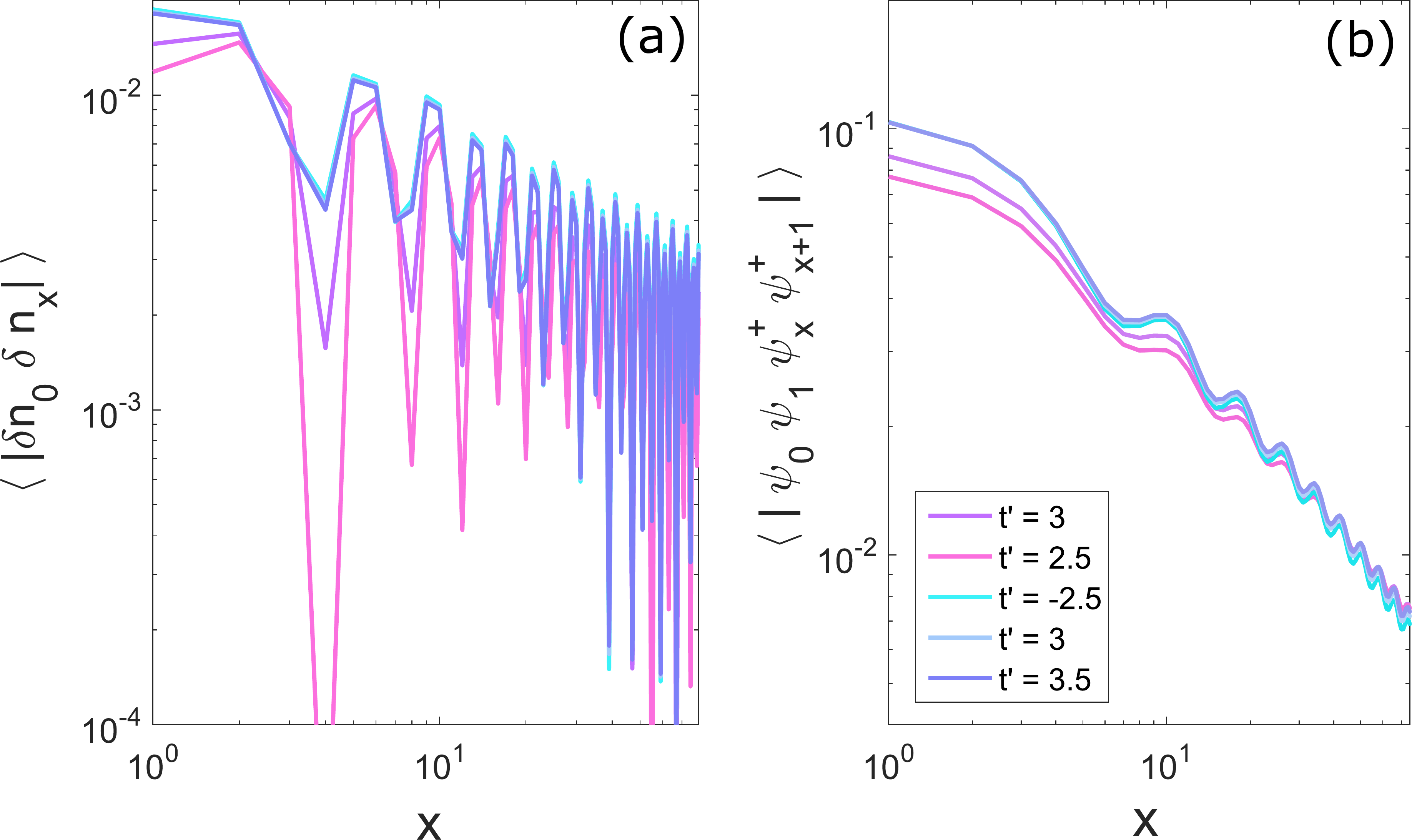} %
    \caption{ (a) density-density correlation function and (b) the pair-pair correlation function for different values of $t'$ in the strong pairing phase. Both correlations decay like a power law indicating that the single particle gap in the strong pairing phase is not due to charge ordering.     }%
    \label{fig:powerlaws} %
\end{figure}

 For completeness we present the density-density and pair-pair correlation functions for different values of $t'$ in the strong pairing phase in Fig. \ref{fig:powerlaws}. We find that both  correlation functions decay like a power law. This rules out the possibility that the gap in the exponential decay of the single-particle correlation function results from charge ordering.

\end{document}